\documentclass[singlecolumn,superscriptaddress,aps,floatfix,prb,10pt]{revtex4-2}
\usepackage[T1]{fontenc}
\usepackage{amssymb,amsmath}
\usepackage{graphicx}
\usepackage{color}
\usepackage{bbm}
\usepackage{bbold}
\usepackage{soul}
\usepackage[dvipsnames]{xcolor}
\usepackage{ulem}
\newcommand{\ah}{\hat{a}}
\newcommand{\sigmah}{\hat{\sigma}}
\newcommand{\Uh}{\hat{U}}
\newcommand{\uh}{\hat{u}}

\newcommand{\ket}[1]{| #1 \rangle}
\newcommand{\bra}[1]{ \langle #1 |}

\begin{document}
\title{Realising and compressing quantum circuits with quantum reservoir computing} 
\author{Sanjib Ghosh}
\email{sanjibghosh87@u.nus.edu}
\affiliation{School of Physical and Mathematical Sciences, Nanyang Technological University 637371, Singapore}
\author{Tanjung Krisnanda}
\affiliation{School of Physical and Mathematical Sciences, Nanyang Technological University 637371, Singapore}
\author{Tomasz Paterek}
\affiliation{School of Physical and Mathematical Sciences, Nanyang Technological University 637371, Singapore}
\affiliation{Institute of Theoretical Physics and Astrophysics, Faculty of Mathematics, Physics and Informatics, University of Gda\'{n}sk, 80-308 Gda\'{n}sk, Poland}
\author{Timothy C. H. Liew}
\email{timothyliew@ntu.edu.sg}
\affiliation{School of Physical and Mathematical Sciences, Nanyang Technological University 637371, Singapore}

\begin{abstract}
\section{Abstract} 
Quantum computers require precise control over parameters and careful engineering of the underlying physical system. In contrast, neural networks have evolved to tolerate imprecision and inhomogeneity. Here, using a reservoir computing architecture we show how a random network of quantum nodes can be used as a robust hardware for quantum computing. Our network architecture induces quantum operations by optimising only a single layer of quantum nodes, a key advantage over the traditional neural networks where many layers of neurons have to be optimised. We demonstrate how a single network can induce different quantum gates, including a universal gate set. Moreover, in the few-qubit regime, we show that sequences of multiple quantum gates in quantum circuits can be compressed with a single operation, potentially reducing the operation {time and} complexity. As the key resource is a random network of nodes, with no specific topology or structure, this architecture is a hardware friendly alternative paradigm for quantum computation.
\end{abstract}
\maketitle

\section{Introduction}
\label{Sec:Introduction}
While conventional computers rely on predetermined algorithms for performing tasks, computers based on artificial neural networks are flexible and can learn from example in analogy to a biological brain. Their resilience allows them to be versatile in applications and adaptive to practical situations. For instance, artificial neural networks are used across disciplines for a multitude of tasks \cite{Jones19,Topol19,Hannun19,Nagy19,Vicentini19,Mehta19,Yoshioka19,Hartmann19, Miscuglio20}, and are capable of extracting features from noisy~\cite{Wong93,Borodinov19} or incomplete data~\cite{Che18,Ding19}, as well as perturbed systems~\cite{Ming19}.

An artificial neural network is a system of interconnected nonlinear nodes capable of modelling complex mapping between input and output data. A given map is formed by carefully adjusting the connection weights between the nodes during a training procedure. While neural networks have been used for various applications, most of them are in the form of softwares implemented in conventional computers. Hardware realizations have been sought, however a major challenge is the control of the large number of connections between the nodes in conventional neural network architectures~\cite{Roy19}.

``Reservoir computing'' refers to an alternative neural network architecture where the connections inside the network are taken to be fixed and random~\cite{Schrauwen07,Lukosevicius12}, thus avoiding the overhead of controlling a large number of connections, while keeping its wide range of applications~\cite{Grigoryeva18,Seoane19}. In this context, the fixed random network is known as the ``reservoir''. As it is easier to engineer a fixed and random network than a well controlled one, reservoir computing has been successfully implemented in a variety of physical systems~\cite{Tanaka19,Nakajima20,Kusumoto19,Ballarini20}. Recently, the reservoir computing concept was brought to the quantum domain~\cite{Markovic20}, using networks of quantum nodes~\cite{Fujii17,Nakajima19} and the performance of specific non-classical tasks~\cite{Ghosh19} including quantum state preparation~\cite{Ghosh19StatePre,Krisnanda21} and tomography~\cite{Ghosh20StateTomo}. While these examples operate with quantum systems, they work with classical data either in the input or output and are far from being quantum computers, which should be able to implement unitary transformations (at least approximately) of a quantum state.

In quantum computing, one of the most commonly used architectures is the quantum circuit model where an arbitrary quantum operation is decomposed with elementary quantum gates~\cite{Nielsen02}. Their realization requires precise engineering~\cite{Almudever17}, which has led to the developments of quantum computers with a limited number of qubits so far~\cite{Google50,IBM20}, while the actual number required for meaningful applications is orders of magnitude higher~\cite{Fowler12}. Although any quantum operation can be in principle obtained by applying gate combinations from a small set of quantum gates (referred to as universal gate set), long gate sequences require long operation time and lead to large errors. Alternatively, many frequently used elementary quantum operations can be obtained directly instead of obtaining them as combinations of gates from a universal gate set. However, realising different types of operations has required different types of interaction between the qubits, leading to more complex engineering.

Here, we introduce a scheme of quantum computing based on a reservoir computing framework. As the main network, we take a set of quantum nodes coupled via random and uncontrolled quantum tunneling. This network would traditionally be referred to as a reservoir within the context of neural networks. However, we will refer to the set of quantum nodes as the ``quantum network'' (QN) with its main feature being that its weight connections are not engineered. Similar networks of qubits were used before in Refs.~\cite{Banchi16,Innocenti20} to realise non-trivial multi-qubit gates, where, unlike the reservoir computing architecture, all network weights were optimized. Let us also point out that our scheme has no relation to techniques of reservoir engineering, as we are not operating with a thermal bath or proposing to engineer the QN~\cite{Poyatos96,Verstraete09,Lin13,Kienzler15}. Our use of the term ``network'' is not supposed to convey a particularly large number of nodes. For efficiency we will work with the smallest possible network. 

\begin{figure}[]
\includegraphics[width=0.6\linewidth]{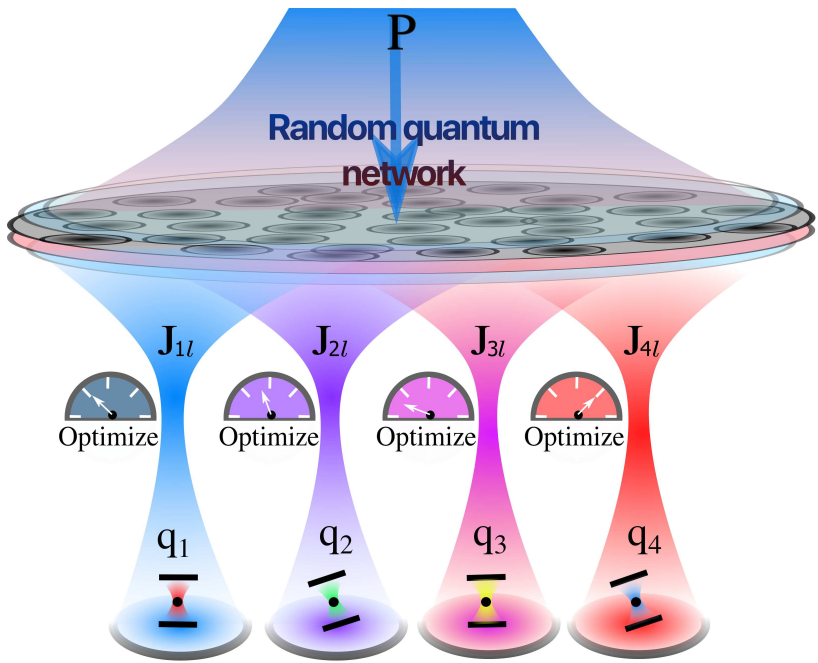}
\caption{The scheme for quantum computing based on a quantum network. The QN is composed of a randomly connected network of nodes (black circles) driven with coherent excitation $P$ (Blue arrow). The qubits on which computation is performed are denoted by $q_k$ ($k$ ranges from $1$ to the number of qubits). Quantum operations on qubits $q_k$ are performed by coupling them through the QN. $J_{kl}$ ($l$ ranges from $1$ to the number of network sites $N$) represent a control layer of tunnelling amplitudes connecting the qubits to the QN. Different colors of qubits correspond to different levels of optimization.} 
\label{TheScheme}
\end{figure}

The QN is connected to a layer of ``computational'' qubits upon which quantum processing is to be performed. We allow the strengths of the weights between the QN and computational qubit layer as the only layer of the network that needs optimization, as illustrated in Fig.~\ref{TheScheme}. We show that a single QN can induce a wide variety of quantum operations on the qubits. In addition to universal quantum gates, we show that our scheme can directly induce non-unitary quantum operations, which is useful to simulate open quantum systems. We emphasise that these are achieved with quantum tunnelling as the only mode of interaction. 

\begin{figure}[]
\includegraphics[width=0.65\linewidth]{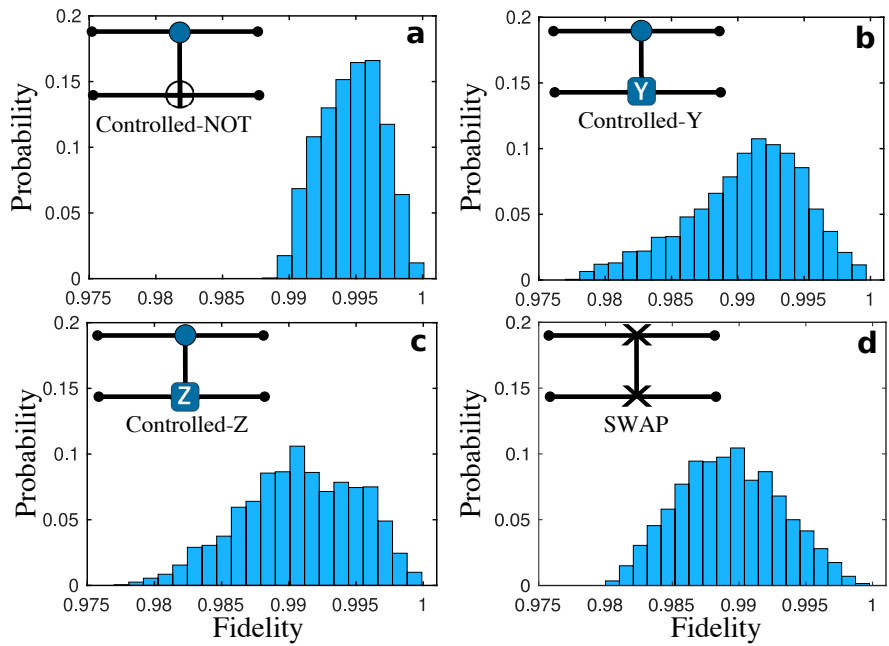}
\caption{Two-qubit gates and their fidelity distributions. \textbf{a}, \textbf{b}, \textbf{c}, and \textbf{d} are the fidelity distributions of the realised controlled-NOT (cNOT), controlled-Y (cY), controlled-Z (cZ) and SWAP gates, as their circuit diagrams are shown in the insets. Here we consider $6$ nodes for the quantum network. The fidelity distributions are obtained over $2000$ uniformly at random generated states. The average fidelities for all gates are larger than $0.99$. Here, we use $E_0/K_0 = (1, 1, 1, 2)$, $P/K_0= (400, 400, 154, 5.5)$ and $\tau K_0/\hbar = (0.23, 0.2, 0.23, 0.3)$ for \textbf{a} to \textbf{d}, respectively and $10$ randomly generated pure states for training. $E_0$, $K_0$, $P$ and $\tau$ are energy, hopping strength, driving strength, and evolution time, respectively.}
\label{TwoQubits}
\end{figure}

\begin{figure}[]
\includegraphics[width=0.95\linewidth]{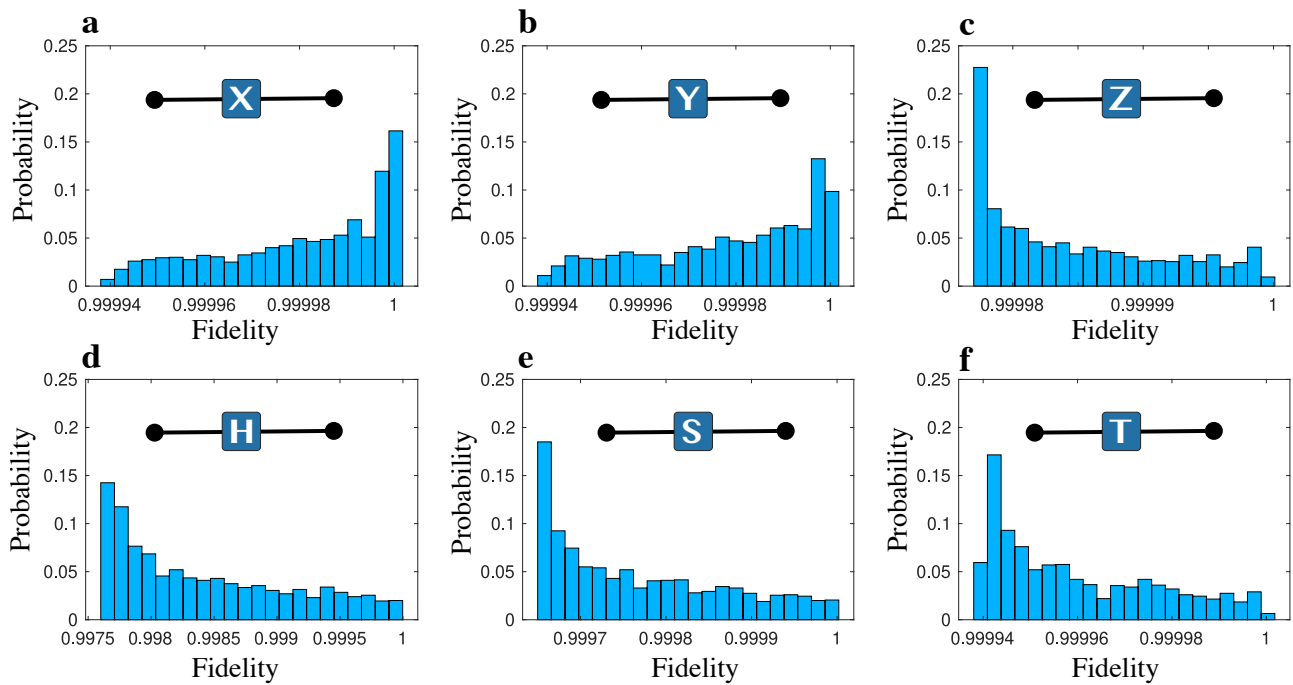}
\caption{Fidelities of the single-qubit gates. We show the fidelity distributions for different single-qubit gates over $2000$ uniformly at random generated states. \textbf{a} to \textbf{f} show the fidelity distributions for Pauli-X, Y, Z, Hadamard, phase and $\pi/8$ gates, respectively. For these gates a single network node is sufficient. A larger quantum network with $6$ sites can also be considered, see Supplementary Section~2. The average fidelities for all the single qubit gates are larger than $0.9984$. Here, we use $P/E_0= (60, 60, 0.1, 4.96, 0.1,0.1)$ and $\tau E_0/\hbar = (3.08, 16.68, 6.28, 1.53, 3.07, 4.71)$ for \textbf{a} to \textbf{f}, respectively and $10$ randomly generated pure states for training. $E_0$, $P$ and $\tau$ are energy, driving strength, and evolution time respectively.}
\label{SingleQubits}
\end{figure}

\begin{figure}[]
\includegraphics[width=0.55\linewidth]{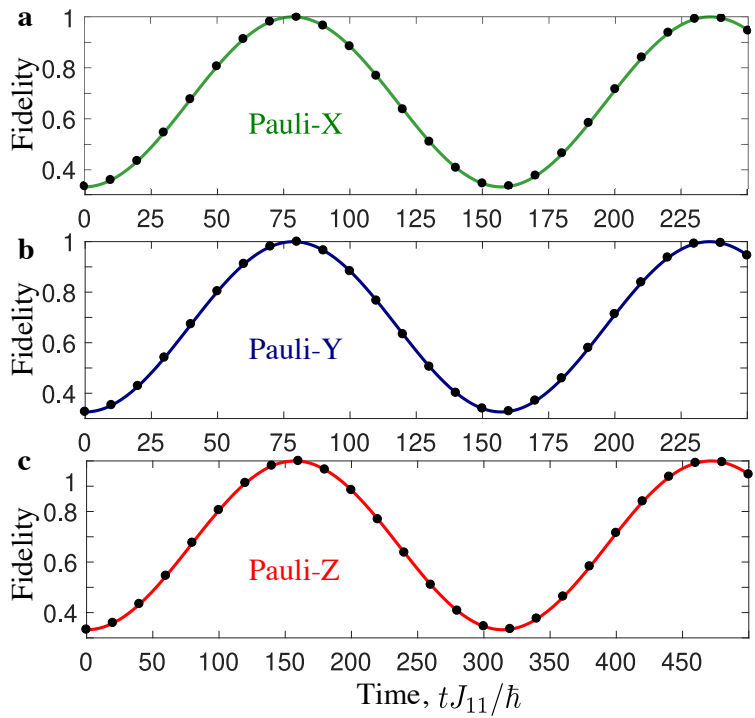}
\caption{A quantum network can induce nonunitary operation. \textbf{a}, Time dependence of qubit purity. \textbf{b}, Distribution of fidelity of a qubit evolving under a Markovian process. We consider a decay strength $\gamma$ and a propagation time $t = 0.5 \hbar/\gamma$ for the Markovian process. Here, we considered a single network node and $2000$ uniformly distributed initial states. $K_0$ is the hopping strength.}
\label{SingleQubitOpen}
\end{figure}

\begin{figure}[]
\includegraphics[width= 0.8 \linewidth]{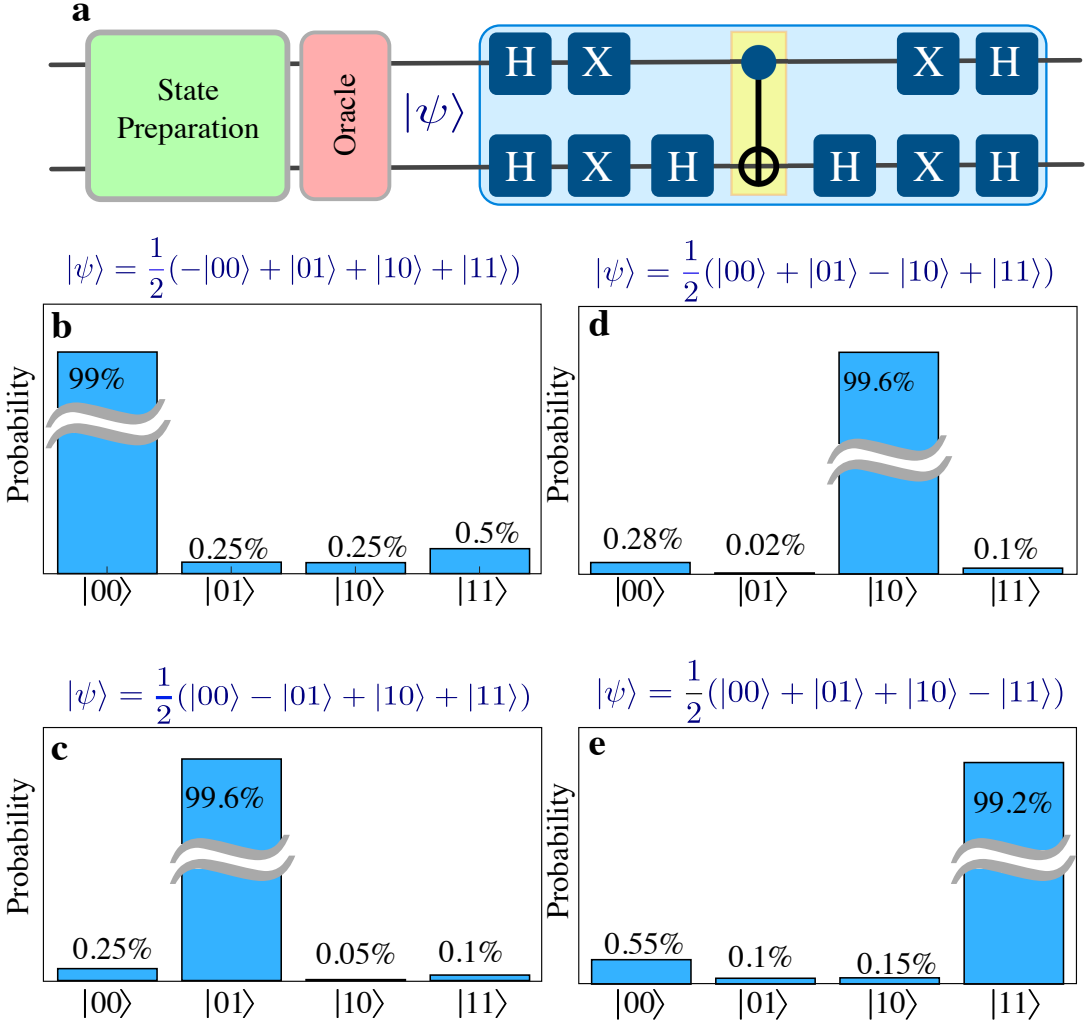}
\caption{Implementation of the two-qubit Grover's algorithm with a quantum network. \textbf{a} A quantum circuit implementing the two-qubit Grover's algorithm. The circuit in the blue box contains $11$ gates, which we have replaced with a single operation acting on a state $|\psi\rangle$. Here, $|\psi\rangle$ is created with a state preparation circuit (green box) and an oracle (red box). We achieved an average fidelity $0.99$ with six sites in the quantum network. The obtained output probabilities (expressed in percentage) for the Grover's search task are presented in panels \textbf{b}-\textbf{e} for different $|\psi\rangle$, where the x-axis represents the measurement basis states. Here, we use $E_0/K_0 = 300 $, $P/K_0= 98 $ and $\tau K_0/\hbar = 10.6$ and $10$ randomly generated pure states for training. Here, H, X and the yellow box represent Hadamard, Pauli-X and controlled-NOT gates, and $E_0$, $K_0$, $P$ and $\tau$ are energy, hopping strength, driving strength, and evolution time, respectively.}
\label{TwoQGrover}
\end{figure}

\begin{figure}[]
\includegraphics[width = 0.8 \linewidth]{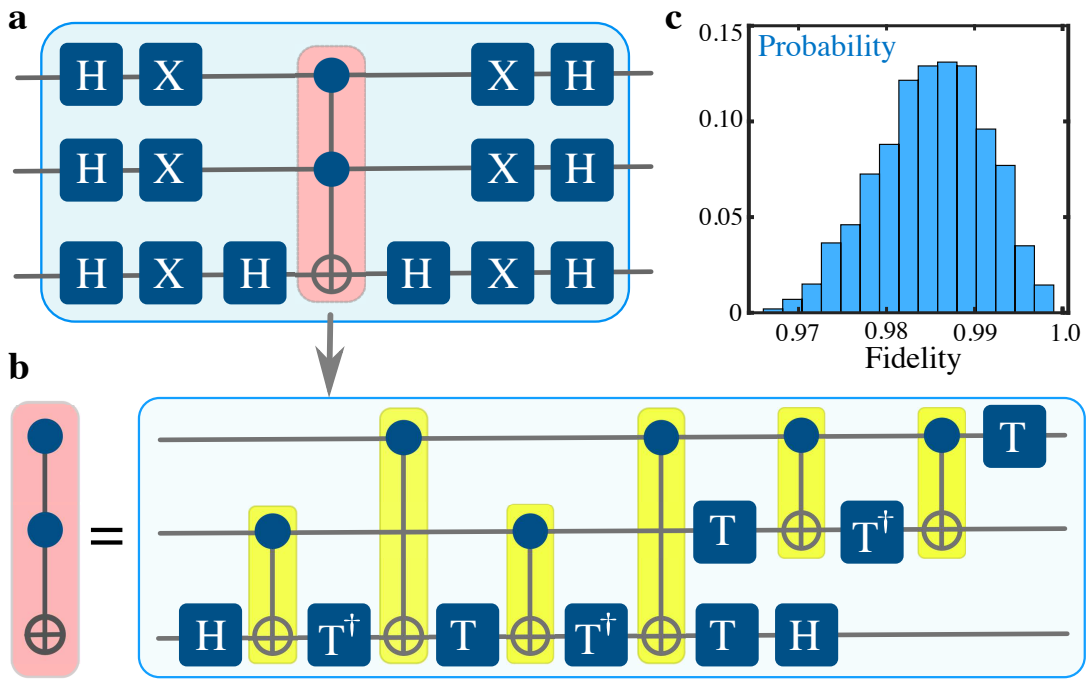}
\caption{Implementation of the three-qubit Grover's algorithm with a quantum network. \textbf{a} A quantum circuit (blue box) implementing the three-qubit Grover's diffusion operator, which includes $29$ single- and two-qubit gates. The circuit in the red box represents a three-qubit Toffoli gate. We have replaced the whole circuit with a single operation implemented by a quantum network (QN). We achieved an average fidelity $0.986$ with five sites in the QN. \textbf{b} Quantum circuit (blue box) composed with single- and two-qubit gates for a three-qubit Toffoli gate (red box) used in \textbf{a}. \textbf{c} The distribution of fidelity for $2000$ randomly generated input states. Here, X, H, T, T$^\dagger$ and yellow boxes represent Pauli-X, Hadamard, $\pi/8$, $-\pi/8$ and controlled-NOT gates, respectively. We have use $E_0/K_0 = 2.99\times 10^3 $, $P/K_0= 994.25 $ and $\tau K_0/\hbar = 1.22$ and $10$ randomly generated pure states for training, where $E_0$, $K_0$, $P$ and $\tau$ are energy, hopping strength, driving strength, and evolution time, respectively.}
\label{ThreeQGrover}
\end{figure}

\section{Result}
A QN is formed with a network of two-level quantum nodes, which are interconnected via quantum tunnelling with random weights and excited with a classical field. The Hamiltonian is
\begin{eqnarray}
\hat{H}_R =\sum_l E_l \ah^\dagger_l \ah_l &+& \sum_{\langle l l' \rangle} K_{l l'} ( \ah_l^\dagger \ah_{l'} + \ah_{l'}^\dagger \ah_l ) \notag \\
&+& \sum_l (P_l^* \ah_l  + P_l \ah_l^\dagger) ,
\label{ReservoirHamil}
\end{eqnarray}
The field operators $\ah_l^\dag$ and $\ah_l$ are the raising and lowering operators of a two-level system, which can be defined as $\ah_l^\dag = |e\rangle \langle g|_l$ and $\ah_l = | g \rangle \langle e |_l$, where $|g \rangle $ and $|e \rangle $ are the ground and excited states of the system with index $l$. $E_l$ and $K_{ll'}$ are site-dependent energies and nearest-neighbour hopping amplitudes, uniformly distributed in the intervals $[\pm E_0/2]$ and $[\pm K_0/2]$, respectively. 
The last term in Eq.~\ref{ReservoirHamil} corresponds to the driving by a classical field. For simplicity we consider a uniform driving, such that $P_l=P$. All calculations are performed with an open boundary condition (i.e., finite lattice with no periodicity) for the QN Hamiltonian.

The QN interacts with a set of computational qubits with the coupling weights $J_{kl}$, such that the whole system is described by the Hamiltonian
\begin{eqnarray}
\hat{H} = \hat{H}_R + \sum_{kl} \left( J^*_{kl} \, \sigmah_k^+ \ah_l +  J_{kl} \, \ah_l^\dagger \sigmah_k^- \right),
\label{Hamiltonian}
\end{eqnarray}
where $\sigmah_k^\pm = \sigmah_k^x \pm i \sigmah_k^y  $ with $\sigmah_k^x$ and $\sigmah_k^y$ being the Pauli-X and Y operators for the qubit $q_k$. Note that the computational qubits do not directly interact with each other, but only via the QN through quantum tunnelling. Our proposition is to induce quantum operations on the qubits $q_k$ by switching on the tunnelling amplitudes $J_{kl}$ for a time $\tau$. The parameters are chosen with a four-step method: (\textbf{a}) We consider a sufficiently large QN with random fixed hopping amplitudes $K_{ll'}$ and energies $E_l$. (\textbf{b}) $P$ and $\tau$ are chosen (but not fine-tuned) in regimes where the output fidelity is high. (\textbf{c}) We then train (fine-tune) the tunnelling amplitudes $J_{kl}$ by maximising the fidelity. (\textbf{d}) For very small QNs, we train $J_{kl}$, $P_l$ and $\tau$ to maximise the fidelity.

For training, we sample a set of pure input states for the qubits and compute fidelity of the states resulting from the QN compared to the ideal states corresponding to a desired quantum operation. The optimisation is performed using a hybrid genetic Nelder-Mead algorithm to set the tunnelling amplitudes $J_{kl}$ (details in Supplementary Section~1). In practice, this supervised learning procedure requires access to a set of ideal input-output state pairs (we considered $10$ randomly generated states), which can either be calculated theoretically or taken as a resource in an experimental setup. We allow $J_{kl}$ to be complex, for generality, however this is not strictly necessary for our scheme. Once $J_{kl}$ are optimised, the fidelity is retested with an independent sample of input states.

Regarding the practical feasibility we note that the Fermi-Hubbard model represented by the Hamiltonian $\hat{H}$ has efficiently been implemented using cold atoms in optical lattices~\cite{Esslinger10,Hofstetter18,Tarruell18} and is expected to be accessible in nonlinear cavity arrays~\cite{Carusotto09,Bardyn12}, depending on the strongly interacting photon regime. Substantial progress has been made toward reaching this regime using a variety of systems, including Rydberg atoms in high quality factor cavities~\cite{Chang14}, photonic crystal structures~\cite{Angelakis17}, superconducting circuits~\cite{Vaneph18}, exciton-polaritons~\cite{Delteil19}, and trion-polaritons~\cite{Emmanuele19, Kyriienko19}. A variety of physical implementations of coupling of quantum emitters to waveguides~\cite{Pierre19} or resonators~\cite{Roy17} have also been considered, where lattices of superconducting qubits~\cite{Fitzpatrick17} have been particularly successful. These classes of systems are typically described by the Jaynes-Cummings-Hubbard model~\cite{Nissen12,Snijders18} where bosonic cavity modes are used to couple separated fermionic modes. The bosonic modes can be eliminated (under some conditions), giving an effective Fermi-Hubbard model~\cite{Scarlino19}.

\subsection{Quantum operations}
Here we describe the protocol for inducing quantum operations on the computational qubits. An operation begins at time $t=0$ with the initialization of qubits. We sample the initial states $\ket{\varphi_\text{in}}$ uniformly at random, assuming that the QN starts in the vacuum state $\ket{\text{vac}}_R$. The whole system is allowed to evolve up to a time $t=\tau$, corresponding to the unitary operator $\Uh = \exp [-i \hat{H}\tau/\hbar]$. The final state of the combined system is thus given by $\ket{\Psi_\text{out} }= \Uh \,\,\ket{\varphi_\text{in}}\otimes \ket{\text{vac}}_R $. The final state of the qubits is given by $\rho_{q}= \text{Tr}_R [\ket{\Psi_\text{out}}\bra{\Psi_\text{out}}]$ where $\text{Tr}_R[\dots]$ represents the partial trace that traces out the QN. For each initial state we compute the fidelity given by the overlap of the ideal final quantum state $\ket{\varphi_\text{ideal}}=\uh_{q} \ket{\varphi_\text{in}}$ and the obtained state $\rho_q$:
\begin{eqnarray}
F = \bra{\varphi_\text{ideal}}\rho_q \ket{\varphi_\text{ideal}}
\label{Fidelity}
\end{eqnarray}
where $\uh_{q}$ is the desired quantum operation for the qubits. We plot fidelity histograms to show that the realised gates are almost perfect for any input state.

We first show that the same QN can realize a range of different two-qubit gates, e.g., controlled inversion (cNOT), controlled-Y (cY), controlled-Z (cZ) and qubit swap (SWAP), see Fig.~\ref{TwoQubits}. A specific gate operation is induced with well chosen tunnelling amplitudes $J_{kl}$ and appropriate values of $P$ and $\tau$. In Fig.~\ref{SingleQubits}, we also demonstrate that high fidelity single-qubit gates are realized with a QN consisting of only one node. The same can be achieved with larger QNs, see Supplementary Section~2. These quantum gates, together with the two-qubit gates, form a universal gate set, from which any quantum operation can be constructed in principle.

Although our demonstrations are based on only few-qubit quantum operations, it is to be noted that, even within this few-qubit regime, many frequently used quantum gates are not native gates in most physical systems. Consequently, these gates are then composed of the native universal gate set. As a result, the depth of the circuits increases. Our idea is that since the same QN can induce a large number of quantum operations, this will reduce the depth (compressing) of the circuits. Our study also opens up the possibility of inducing quantum operations involving large number of qubits with a single QN, however that requires further study for a definitive answer.

\subsection{Nonunitary operations}
In general, the operations induced by the QN on the qubits can be nonunitary. The signature of the nonunitary nature of the operation can be observed in the purity of the qubits, shown in Fig.~\ref{SingleQubitOpen}\textbf{a}. The purity oscillates and reaches $1$ only at certain times (only at which the effective operation on the qubit can be considered unitary). We conclude that the system can also be trained to perform non-unitary gates. For a demonstration we consider a Markovian dynamics for the qubit given by the master equation: $\hbar\dot{\rho} = (\gamma/2) (2\sigma^- \rho \sigma^+ - \sigma^+ \sigma^- \rho- \rho \sigma^+\sigma^- )$. Using the QN, we obtain a non-unitary quantum operation equivalent to the same induced by the master equation of the qubit (see Fig.~\ref{SingleQubitOpen} \textbf{b}).

\subsection{Compression of quantum circuits}
We have achieved a universal gate set from single- and two-qubit quantum gates with the QN. In the quantum circuit model, an arbitrary unitary is approximated with a sequence of single- and two-qubit gates. Using the same method, we achieve universality of a QN. Here, a single QN induces different quantum gates in a quantum circuit by changing the output tunnelling amplitudes $J_{kl}$. However, a QN has an important advantage. A long sequence of quantum gates in a circuit model can be replaced with a single operation by a QN. For instance, a two-qubit Grover's algorithm~\cite{Grover96,Brickman05} can be implemented in one step with a QN, while the circuit model requires several quantum gates in sequence (see Fig.~\ref{TwoQGrover}).

In Fig.~\ref{ThreeQGrover}, we present numerical evidence for the possibility of replacing an entire quantum circuit by a single operation. A diffusion operator for the three-qubit Grover's algorithm is shown in Fig.~\ref{ThreeQGrover}\textbf{a}. This circuit includes a three-qubit Toffoli gate, which can also be implemented with single- and two-qubit gates, see Fig.~\ref{ThreeQGrover}\textbf{b}. The whole circuit would require 29 quantum gates in the conventional quantum circuit architecture. In contrast, a QN can implement all the gates by a single operation. The distribution of the output fidelity is plotted in Fig.~\ref{ThreeQGrover}\textbf{c}. We note that although considering fault-tolerance in Grover's algorithm, as it is the case in the circuit model, could be interesting, it is beyond the scope of this paper. 

For implementing Grover's search algorithm, a diffusive operator $\hat{ \mathcal{D}}$ is required to operate $\sqrt{N}$ times on an $N$-qubit initial state to find out the search query. The operator $\hat{ \mathcal{D}}$ can be expressed as,
 \begin{eqnarray}
\hat{ \mathcal{D}} = H^{\otimes N} \left( \openone- 2 \ket{0^{\otimes N}} \bra{0^{\otimes N}} \right) H^{\otimes N}
\end{eqnarray}
where $H^{\otimes N}$ stands for Hadamard gates applied on all $N$ qubits and $\ket{0^{\otimes N}}$ represents the qubits in their ground state. For $N=2$ and $3$, the quantum circuits composed of single- and two-qubit gates are presented in Fig.~\ref{TwoQGrover} and Fig.~\ref{ThreeQGrover}, respectively.

\subsection{Robustness} 
Our scheme is inspired by reservoir computing frameworks, which are known for their robustness against imprecisions in the fabrication of the network. Similarly here quantum operations can withstand any undesirable change in the QN, provided that we retrain the amplitudes $J_{kl}$. Indeed, as shown in the Supplementary Section~3, we find that the use of a neural network approach means that the system automatically learns how to deal with static errors in fabrication.

\subsection{Physical systems}
In our considered QN, quantum tunnelling is the only mode of interaction between the nodes. It is thus in principle compatible with a wide range of experimental platforms, including essentially all platforms currently considered for quantum computing, but with far less stringent requirements on the control of coupling between nodes. For example, let us mention ultracold atoms or ions (arguably the most advanced platform), cavity quantum electrodynamic systems (which enjoy relatively accessible measurement via optics), circuit quantum electrodynamic systems (which can be considered more accessible in setup)~\cite{Blais20}, or novel platforms such as those based on the internal states of molecules~\cite{Gaita-Arino19}, coupled Bose-Einstein condensates~\cite{Byrnes12} and lattices of exciton-polaritons~\cite{Boulier20}. In some systems, interactions such as the Coulomb interaction are present in addition to tunnelling, which can in principle further enrich the complexity of quantum networks and potentially improve their computing capacity. Our scheme is particularly applicable for systems like quantum dots (QDs) where positioning  them in regular patterns is a challenge. Consequently, these systems face challenges with disorder in the couplings between QDs, but this poses no problem for our scheme based on random coupling between nodes. The reduction in the number of gates may help less developed systems to implement examples of complete algorithms, even without having the relatively larger number of coupled qubits available in advanced ultracold ion traps.

\section{Conclusion}
We have presented a platform for quantum computing where an underlying set of quantum nodes connects computing qubits and a learning algorithm is used to adapt the system to a particular quantum operation. Several previous works have considered how quantum neural networks can enhance the efficiency of solving classical tasks~\cite{Fujii17}, while others have considered the use of assumed quantum computers~\cite{Cong19,Havlicek19} and quantum annealers~\cite{Amin18} in neuromorphic architectures. In contrast, here we imagine a neuromorphic architecture that can allow a set of quantum nodes to realize quantum computation. The learning of quantum operations from quantum networks has been considered before, based on nonlocal spin coupling~\cite{Banchi16} and adiabatic pulse control~\cite{Zahedinejad16}. The advantage of the QN architecture introduced here is that only quantum tunnelling is considered for network connections, which is readily accessible in many systems (e.g. photonics, polaritons, cold atoms, and trapped ions), and that only a small subset of total network connection weights need to be controlled. We note that a very simple learning algorithm in the optimization of the tunnelling amplitudes was used. The application of more advanced evolutionary algorithms in quantum control would likely lead to improved results for our system~\cite{Yang19}. Alternatively, emerging quantum assisted algorithms~\cite{Khatri19} can be used for operations with large number qubits, for instance, using hybrid quantum-classical algorithms based on gradient descent, since gradients can be efficiently measured on near term hardwares~\cite{Banchi21}. We note that while an advantage of quantum reservoir computing is that effects from disorder are corrected for at the training stage, a limitation is that each individual device would require separate training. This could make large-scale deployment by a manufacturer difficult.

\section{Author's Contribution}
SG conceived the project through discussion with TCHL, TP, and TK. SG performed all the calculations with helps from TK. All authors wrote the paper, discussed the results, and agreed with the conclusions. TCHL supervised the project.

\section{Acknowledgements}
SG, TK and TL were supported by the Ministry of Education (Singapore), grant No. MOE2019-T2-1-004. TP acknowledges the Polish National Agency for Academic Exchange NAWA Project No. PPN/PPO/2018/1/00007/U/00001.

\section{References}

\section*{Supplementary Information}

\section{Supplementary Section 1: Training}
One of the main features of neural network frameworks is that they learn from examples without being told the specific rules of the task. Here we take advantage of this powerful feature and use it for realizing various quantum operations by training the QN. While in conventional approaches, different quantum gate operations require realizing different types of interactions between the qubits, here, the same quantum network is used to obtain different operations. Given a quantum operation $\uh_q$, which we want to realize, we consider a set of example quantum states $\ket{\varphi^{(j)}_\text{in}}$ for the qubits for training. For each example quantum state $\ket{\varphi^{(j)}_\text{in}}$, we calculate the fidelity $F_j$ given by Eq.~3 of the main text. We numerically maximize the average fidelity
\begin{eqnarray}
\bar{F} = \frac{1}{N}\sum_{j=1}^NF_j
\end{eqnarray}
with optimal choice of tunnelling amplitudes $J_{kl}$, where $N$ is the number of quantum states in the training set. We empirically find that for two-qubit gate operation $10$ randomly generated quantum states are sufficient for training.

Training of the quantum neural network is an optimization process. The average fidelity $\bar{F}(J_{kl})$ is a nonlinear function of the tunnelling amplitudes $J_{kl}$. For a small number of parameters $J_{kl}$, a deterministic method such as the Nelder-Mead simplex algorithm is sufficient to achieve the optimum condition. However, for large numbers of $J_{kl}$, we use a stochastic genetic algorithm to find the optimum condition starting from a set of initial guesses, which approach the optimum point in a random process.

Genetic algorithms are inspired by the biological evolution based on natural selection. The process starts with a random set of populations that goes through a natural selection procedure based on a fitness criterion. The fittest individuals reproduce the next generation of populations through a cross breeding procedure. Random mutation in the new generation ensures diversity among the populations.

Here, $J_{kl}$ with all $k$ and $l$ represent one individual in a genetic algorithm. We define a population with a set of $M$ such individuals $J_{kl}^m$ where $m=1,2\dots M$. We start with a random choice of population. The fitness criterion is defined through calculating the average fidelity $\bar{F}(J^m_{kl})$ for an individual $m$.

The next generation of population is reproduced by the two individuals with largest average fidelities, keeping the total population size fixed to $M$. The next generation $J_{kl}^{'m}$ is born with the cross breeding rule:
\begin{eqnarray}
J_{kl}^{'m}  = (J_{kl}^{p} +   J_{kl}^{q})/2 + \delta f_\text{ran}
\end{eqnarray}
where $\delta$ and $ f_\text{ran}$ represent a mutation rate and a Gaussian random number, respectively. The process of natural selection is then repeated for the new generation until the optimum condition is found.

\section{Supplementary Section 2: Realising single-qubit gates with $6$ reservoir sites}
{In the main text, we have shown that single-qubit gates can be obtained with a single-site QN. However, we can obtain these gates with a QN of 6 sites. Since one site in the reservoir is sufficient, we can choose only one tunnelling amplitude $J_{11}$ to be non-zero (all others are set to zero) when considering a larger QN. With such a consideration we indeed obtained high fidelity single-qubit gates, see Supplementary Figure~\ref{SingleQubitGates6QN}.}
\begin{figure}[]
\centering
\includegraphics[width=.9\textwidth]{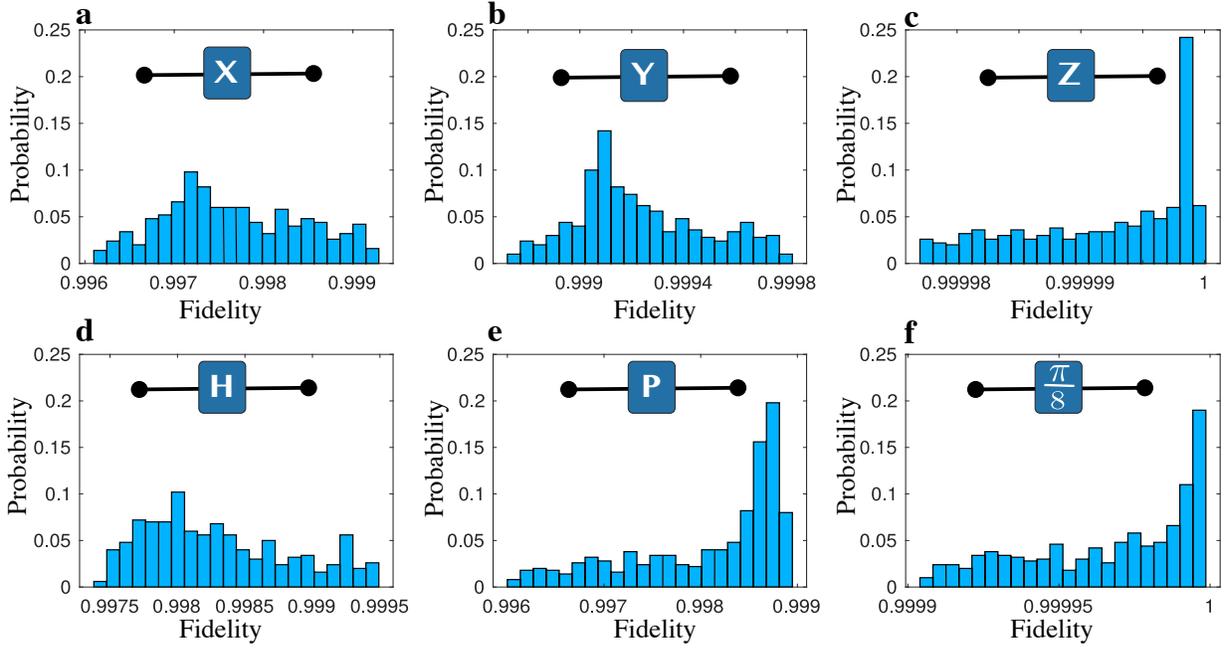}
\caption{Single qubit gates obtained with $6$ sites in a quantum network. \textbf{a} to \textbf{f} show the fidelity distributions for Pauli-X, Y, Z, Hadamard, phase and $\pi/8$ gates, respectively. Here, we use $E_0/K_0 = (1, 1, 1, 1,1,1)$, $P/K_0= (60, 60, 0.1, 50, 0.1, 0.1)$ and $\tau K_0/\hbar = ( 0.89, 0.1, 0.01, 0.15, 12.33, 13.75)$ for \textbf{a} to \textbf{f}, respectively and $10$ randomly generated pure states for training. $E_0$, $K_0$, $P$ and $\tau$ are energy, hopping strength, driving strength, and evolution time, respectively.} 
\label{SingleQubitGates6QN}
\end{figure}

\section{Supplementary Section 3: Robustness}
In our scheme, a quantum operation is learned to be performed with a random network. Thus, as is the case in classical reservoir computing, we expect robustness against fabrication imprecisions in the QN itself, where the parameters $E_l$ and $K_{ll'}$ are taken random. For a given QN, the quantum tunnelling amplitudes $J_{kl}$ are trained to best learn a quantum operation. If the network parameters are changed, we can consider it as a different network, and will retrain $J_{kl}$ to ensure the robustness of the quantum operation. This means that no precision engineering is required for the QN itself. This is major advantage of this computing framework compared to the traditional neural networks, where all the connections in a network need to be precisely chosen.

\begin{figure}[h]
\includegraphics[width=0.3\textwidth]{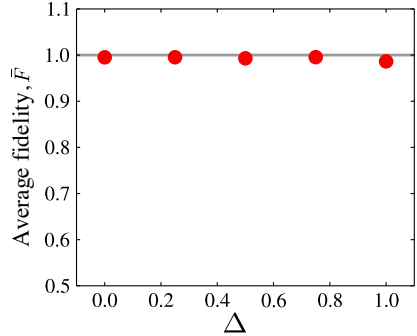}
\caption{Average fidelity $\bar{F}$ for controlled-NOT gate induced by a QN with random parameters as a function of $\Delta$. Here $\Delta$ is the strength of randomness.}
\label{RobustnessCheckCNOT}
\end{figure}

For demonstration, let us consider a QN with the parameters $E_l$ and $K_{ll'}$ induces the CNOT operation. Let us then change them randomly with a strength $\Delta $, such that $E_l \to E_l + \Delta r_l$ and $K_{ll'} \to K_{ll'} + \Delta r_{ll'}$, where $r_{l}$ and $r_{ll'}$ are random variables drawn from the interval $[-1, +1]$. We retrain $J_{kl}$ for each set of parameters. We then calculate the average fidelity for each of the sets. Supplementary Figure~\ref{RobustnessCheckCNOT} shows indeed the average fidelity is little affected by the random change in the network parameters. Note that for $\Delta\ge0.5$, the randomness is as high as the values of the parameters $E_l$ and $K_{ll'}$.

\section{Supplementary Section 4: One-qubit gates}
Here, we consider a special case where one input qubit interacts with a one-site QN. The QN has an onsite energy $E_1$, a coherent driving (pump) $P$, and negligible decay. The dynamics of the whole system is unitary, where the Hamiltonian is written as
\begin{equation}
\label{EQ_H1qgates}
H=E_1\hat a_1^{\dagger}\hat a_1+P\hat a_1^{\dagger}+P^*\hat a_1+J_{11}(\hat \sigma^+_1 \hat a_1+\hat a_1^{\dagger} \hat \sigma^-_1),
\end{equation}
with $J_{11}$ being the strength of the coupling term. Note that we have used $\hat \sigma^-_1$ ($\hat a_1$) as the annihilation operator of the input qubit (the QN site). Below we will present the working parameters for the dynamics of this system such that the evolution of the input qubit implements the X, Y, and Z gates given by the Pauli operators
\begin{equation}
\hat \sigma^x=\left( \begin{array}{cc}
0 & 1  \\
1 & 0 \\
\end{array} \right),\:\:
\hat \sigma^y=\left( \begin{array}{cc}
0 & -i  \\
i & 0 \\
\end{array} \right),\:\:
\hat \sigma^z=\left( \begin{array}{cc}
1 & 0  \\
0 & -1 \\
\end{array} \right),
\end{equation}
respectively. We will present two regimes, namely the low and high energy limit, in which one can implement the one-qubit gates on the input qubit.

\section{Supplementary Section 5: Two-qubit gates (via direct interactions)}
Consider two interacting two-level systems as in the effective picture discussed in the main text. Each qubit has an onsite energy $E_j$, a coherent pumping $P_j$, and no decay. Together with a hopping interaction term, one writes the Hamiltonian as
\begin{align}
H=\sum_{j=\{1,2\}} E_j\hat a_j^{\dagger}\hat a_j+P_j\hat a_j^{\dagger}+P_j^*\hat a_j 
+J(\hat a_1 \hat a_2^{\dagger}+\hat a_2 \hat a_1^{\dagger}).
\end{align}
It will be shown below that this type of system is able to implement universal two-qubit gates on its initial state.
We will start by presenting the working parameters for important gates, such as the square-root-swap (sSWAP), control-X (cNOT), control-Y (cY), control-Z (cZ), square-root-iSwap (siSWAP), and swap (SWAP).
These gates are written as:
\begin{align}
\hat U_{\text{sSWAP}}&=\frac{1}{2}\left( \begin{array}{cccc}
2 & 0 &0 &0 \\
0 & (1+i) &(1-i) &0 \\
0 & (1-i) &(1+i) &0\\
0 & 0 &0 &2 \\
\end{array} \right),\\
\hat U_{\text{cNOT}}&=\left( \begin{array}{cccc}
1 & 0 &0 &0 \\
0 & 1 &0 &0 \\
0 & 0 &0 &1 \\
0 & 0 &1 &0 \\
\end{array} \right), \\
\hat U_{\text{cY}}&=\left( \begin{array}{cccc}
1 & 0 &0 &0 \\
0 & 1 &0 &0 \\
0 & 0 &0 &-i \\
0 & 0 &i &0 \\
\end{array} \right), \\
\hat U_{\text{cZ}}&=\left( \begin{array}{cccc}
1 & 0 &0 &0 \\
0 & 1 &0 &0 \\
0 & 0 &1 &0 \\
0 & 0 &0 &-1 \\
\end{array} \right), \\
\hat U_{\text{siSWAP}}&=\frac{1}{\sqrt{2}}\left( \begin{array}{cccc}
\sqrt{2} & 0 &0 &0 \\
0 & 1 &i &0 \\
0 & i &1 &0\\
0 & 0 &0 &\sqrt{2} \\
\end{array} \right), \\
\hat U_{\text{SWAP}}&=\left( \begin{array}{cccc}
1 & 0 &0 &0 \\
0 & 0 &1 &0 \\
0 & 1 &0 &0\\
0 & 0 &0 &1 \\
\end{array} \right).
\end{align}

\subsection{Supplementary Section 6: Simulations}
\begin{figure}[h]
\includegraphics[width=0.5\textwidth]{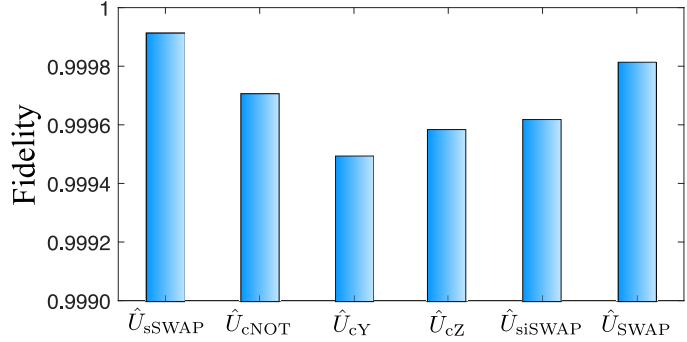}
\caption{Fidelity $\bar F$ of various two-qubit gates for the parameters listed in Supplementary Table \ref{TB_2qpara}.}
\label{FIG_2qgates}
\end{figure}

We performed a search algorithm realising the above two-qubit gates. Our numerical results show that all these gates can be achieved with gate fidelity $ \bar F>0.999$ (see Supplementary Table \ref{TB_2qpara} for exemplary parameters). Supplementary Figure~\ref{FIG_2qgates} shows a comparison of the high fidelity achieved with the search method for different gates.
Note that genetic algorithms have been used for realising $\hat U_{\text{sSWAP}}$, $\hat U_{\text{cNOT}}$, $\hat U_{\text{cY}}$, and $\hat U_{\text{cZ}}$.
The parameters for $\hat U_{\text{siSWAP}}$ and $\hat U_{\text{sSWAP}}$ are not sensitive to small changes and therefore we did not perform precise search algorithms.

\begin{table}[!t]
\begin{center}
\caption{Exemplary parameters for the realisation of two-qubit gates. We have used $\tau=E_2t/\hbar$. All the gates are achieved with $\bar F >0.999$ with $N=10^5$. Note that the parameters for the $\hat U_{\text{siSWAP}}$ and $\hat U_{\text{SWAP}}$ gates are not as sensitive as the others. Especially for $\hat U_{\text{siSWAP}}$, one gets very high fidelity in the limit $E_1=E_2\gg P_1=P_2,J$.
}\label{TB_2qpara}
\smallskip
\begin{tabular}{|l|c|c|c|c|c|}
\hline
Two-qubit &  \multicolumn{5}{c|}{Parameters}\\
\cline{2-6}
Gates & $E_1/E_2$ & $P_1/E_2$ & $P_2/E_2 $& $J/E_2$ & $\tau$ \\
\hline
\hline
$\hat U_{\text{sSWAP}}$ & $0.923091$ & $5.158696$ & $5.155802$ & $0.937275$ & $48.168039$\\
\hline
$\hat U_{\text{cNOT}}$ & $140.703597$  & $0.958346$ & $140.627941$ & $2.826258$& $40.303524$ \\
\hline
$\hat U_{\text{cY}}$ & $138.217022$  & $-0.089054-0.920161i$ & $0.079873-138.295970i$ & $2.881602$ & $40.778441$ \\
\hline
$\hat U_{\text{cZ}}$ & $1.006724$  & $1.094922$ & $0.932635$ & $117.958714$ & $45.402586$\\
\hline
$\hat U_{\text{siSWAP}}$  & $1$ & $0.01$ & $0.01$ & $0.01$ & $1181$\\
\hline
$\hat U_{\text{SWAP}}$  & $1$ & $1.5$ & $1.5$ & $42.8$ & $37.6$\\
\hline
\end{tabular}
\end{center}
\end{table}

\subsection{Supplementary Section 7: Universal quantum gates}
We note that each of the two-qubit gates presented above, apart from the SWAP gate, combined with single-qubit gates form a universal set. In this way, they are equivalent to each other. As an example, we recall that the cNOT gate can be created, by the cY, cZ, sSWAP or siSWAP gate with the help of single qubit gates as follows
\begin{align}
\hat U_{\text{cNOT}}&=[\openone\otimes \hat R_{z}(-\pi/2)] \:\hat U_{\text{cY}}\: [\openone \otimes \hat R_z(\pi/2)],  \notag\\
\hat U_{\text{cNOT}}&=[\openone\otimes \hat R_{y}(\pi/2)] \:\hat U_{\text{cZ}}\: [\openone \otimes \hat R_y(-\pi/2)], \notag \\
\hat U_{\text{cNOT}}&=[\openone\otimes \hat R_{y}(-\pi/2)]  \: \hat U_{\text{sSWAP}} \: [\hat Z\otimes \openone] \:\notag\\
&\hspace{10mm} \hat U_{\text{sSWAP}}
 \: [\hat R_z(-\pi/2)\otimes \hat R_z(-\pi/2)]\: \notag\\
&\hspace{10mm}[\openone \otimes \hat R_{y}(\pi/2)],\nonumber\\
\hat U_{\text{cNOT}}&=[\hat X\otimes \hat X]\:[\hat R_{y}(-\pi/2)\otimes \openone][\hat R_{x}(\pi/2)\otimes \hat R_{x}(-\pi/2)]\: \notag\\
&\hspace{10mm}\hat U_{\text{siSWAP}}\:[\hat R_{x}(\pi)\otimes \openone]\:\hat U_{\text{siSWAP}}\notag\\
&\hspace{10mm}\: [\hat R_{y}(\pi/2)\otimes \openone]\:[\hat Z\otimes \openone]\:[\hat X\otimes \hat X] \:e^{i\pi/4}.
\end{align}
Note that we have used the following single-qubit rotation matrices.
\begin{eqnarray}
&&\hat R_x(\alpha)=\left( \begin{array}{cc}
\cos(\frac{\alpha}{2}) & -i\sin(\frac{\alpha}{2})  \\
-i\sin(\frac{\alpha}{2}) & \cos(\frac{\alpha}{2}) \\
\end{array} \right), \notag \\
&&\hat R_z(\delta)=\left( \begin{array}{cc}
1 & 0  \\
0 & e^{i\delta} \\
\end{array} \right), \notag \\
&&\hat R_y(\beta)=\left( \begin{array}{cc}
\cos(\frac{\beta}{2}) & -\sin(\frac{\beta}{2})  \\
\sin(\frac{\beta}{2}) & \cos(\frac{\beta}{2}) \\
\end{array} \right).
\end{eqnarray}
Even though different gates can be constructed from specific combinations of a universal set of gates, it should be noted that the ability to directly construct the gate needed for a particular application will bring the highest efficiency in terms of operation time. Indeed the ability of the QN to learn to perform a whole range of different quantum gates is one of its key advantages.

\end{document}